\documentclass[aps,pra,nofootinbib,reprint,twocolumn]{revtex4}

\usepackage{epsfig,amssymb,amsmath}

\usepackage{color}

\def\comment#1{}
\def\labell#1{\label{#1}}
\def\>{\rangle}\def\<{\langle}
\def\togli#1{}

\begin{document}
%\fbox{{\scriptsize Eprint: quant-ph/}}
%\fbox{{\scriptsize Internal report, Lorenzo, \today}}
%\fbox{{\scriptsize Preliminary draft.}}
%\fbox{{\scriptsize Submitted paper.}}

%\fbox{{\scriptsize Preliminary draft.}}
%Title of paper

\title{The Pauli objection} \author{Juan Leon$^{1}$, Lorenzo
  Maccone$^{2}$} \affiliation{\vbox{$^1$Instituto de F\'isica
    Fundamental, CSIC,
    Serrano 113-B, 28006, Madrid, Spain}\\
  \vbox{$^2$Dip.~Fisica and INFN Sez.~Pavia, University~of Pavia, via
    Bassi 6, I-27100 Pavia, Italy}}
\begin{abstract}
  Schr\"odinger's equation says that the Hamiltonian is the generator
  of time translations. This seems to imply that any reasonable
  definition of time operator must be conjugate to the Hamiltonian.
  Then both time and energy must have the same spectrum since
  conjugate operators are unitarily equivalent.  Clearly this is not
  always true: normal Hamiltonians have lower bounded spectrum and
  often only have discrete eigenvalues, whereas we typically desire
  that time can take any real value.  Pauli concluded that
  constructing a general a time operator is impossible (although
  clearly it can be done in specific cases).  Here we show how the
  Pauli argument fails when one uses an external system (a ``clock'')
  to track time, so that time arises as correlations between the
  system and the clock (conditional probability amplitudes framework).
  In this case, the time operator is not conjugate to the system
  Hamiltonian, but its eigenvalues still satisfy the Schr\"odinger
  equation for arbitrary Hamiltonians.
\end{abstract}

%\pacs{03.65.Ta,06.30.Ft,03.65.Ud,03.67.-a}

%Time, measurement of, 06.30.Ft
%Spacetime topology of, 04.20.Gz
%Measurement theory (quantum mechanics), 03.65.Ta
%Quantum information, 03.67.-a
%Entanglement and quantum nonlocality, 
%Measurement theory (quantum mechanics), 03.65.Ta
%quantum mechanics, 03.65.Ta

\maketitle

There are many different ways in which a time operator
\cite{kuchar,anderson,problemoftime} can be introduced into quantum
mechanics. These differences reflect the different physical meanings
that ``time'' may have. We recall the main ones (the following list
is, by necessity, incomplete and clearly the following categorizations
are not clear-cut): (1)~Typical time operators
\cite{popescu,werner,mielnik,delgado,galap} represent a ``time of
arrival'', whose measurement represents the time at which a system is
in a certain location. This is dual \cite{juan} to the Newton-Wigner \cite{nw}
position operator whose measurement represents the position at which a
system is located at a certain time\comment{Check this: is it true
  that the Newton-Wigner position operator represents the position of
  a particle at a given time? It {\em must} be so, but I haven't seen
  it stated explicitly anywhere. Perhaps it's obvious, perhaps
  not. Juan confirmed this in his email, see the introduction of \cite{juan}};
(2)~coordinate time \cite{freden,olkhovski,olkhovski1,caves}; (3)~an
arbitrary parameter (also reinterpreted as ``Newtonian absolute
time'') \cite{stu,fanchi}; (4)~a dynamical variable that parametrizes
different Hilbert spaces \cite{ahar,farhi,wilcz}; (5)~a classical
parameter that cannot be quantized \cite{peresbook,pauli,milburn,qspeed}; (6)~a
parameter that can be quantized, but not using self-adjoint operators
(observables) \cite{holevo,holevo1,freden,olkhovski}; (6)~proper time
\cite{greenbergerproper}; (7) clock time\togli{\footnote{We distinguish
  between proper and clock time because any clock, which measures the
  proper time for a co-moving observer, can be used to define
  coordinate time. We can then use two clocks, one to define
  coordinate time and a co-moving one for proper time.}}
\cite{saleckerwigner,peressaleckerwigner,paw,page,wootters,ak,nostro,morse,zeht,vedral,papersinvedral,rovelli,rovt,rovellibook,gambinipullin,montevideo,esperimentotorino,hilgevoord}.

Here will be dealing with the clock time, mainly focusing on the
Page-Wootters and Aharanov-Kaufherr (PWAK) approach
\cite{paw,page,wootters,ak,nostro}. In this framework, time is defined
as ``what is shown on a clock'', where a clock is some (external)
physical system that is taken as a time reference. Then, the
measurement of time acts as a conditioning that outputs the position
in time of some event that is being gauged by the clock: namely the
emphasis is on the {\em correlations} between a system and the clock
as in `the state of a system {\em given that} the clock shows $t$'.
As shown in \cite{morse,paw,page,wootters,ak,nostro} these
correlations manifest themselves as a ``static'' entangled global
state that satisfies a Wheeler-de Witt \cite{wdw} equation. The PWAK
approach is briefly reviewed in Sec.~\ref{s:paw}.

The Pauli objection \cite{pauli} essentially states that since the
energy is the generator of (continuous) time translations, any time
operator must be conjugate to an energy operator (Hamiltonian) that
has unbounded continuous spectrum, properties which are not satisfied
by the Hamiltonian of typical systems. \comment{Horwitz in
  \cite{timeinterf},pg.735 points out that also Dirac had a similar
  argument in the first edition of his book (which I don't have and
  couldn't find on the internet) and not in the following editions.
  However, \cite{hilgevoord},pg.40 seems to be saying something
  completely different about the first edition. Juan points out some
  relevant literature of the period in his email of 26/7.}The standard
textbook answer to this objection is that time in quantum mechanics
cannot be represented by an operator and is a parameter, external to
the theory, e.g.~\cite{peresbook,zeht}. A different, but connected
objection was put forth by A.~Peres \cite{peresobj}: if the
Hamiltonian is the generator of time translations and the momentum is
the generator of space translations, then the Hamiltonian and the
momentum must always commute, since space and time are independent
degrees of freedom. In this paper we show how the PWAK formalism can
easily bypass these objections and provide an acceptable time
operator.

The main idea is simple: the global Hamiltonian must contain both the
system Hamiltonian (which may have arbitrary spectrum) {\em and} the
clock Hamiltonian, which for an ideal clock must have unbounded
continuous spectrum (physical clocks can clearly only approximate this
ideal situation). It is the clock Hamiltonian that is conjugate to the
time operator, whereas it commutes with the system Hamiltonian which
acts on a different Hilbert space. Then {\em the clock Hamiltonian is
  the generator of clock shifts, hence of ``time'' translations},
whereas the system Hamiltonian is the generator of translations only
of the system state, and not of time. Then $[\hat {T},H]=0$, so that
$H$ and $\hat T$ do not need to have the same spectrum. To overcome
Peres' objection, one notes that the Hamiltonian indeed does commute
with the momentum {\em of the clock}.

There are some arguments whose most extreme formulation says that in
quantum theory ``time is not a quantity at all'' \cite{halvorson},
i.e., there is no way to attribute values to time, be it an operator
or a parameter indistinctly. All these arguments assume that the
spectrum of the Hamiltonian generating the time evolution in the state
space is bounded from below. Indeed Halvorson ruled out the existence
of subspaces of states $s({t}_1 , {t}_2 )$ that can be associated to
time intervals $({t}_1 , {t}_2 )$ and hence, dispensing with the
traditional notion of the passage of time in quantum theory. He
concretely derived \cite{halvorson} the contradiction that, as a
consequence of the Hegerfeldt theorem \cite{hegerfeldt}, $\forall |v\>
\in s({t}_1 , {t}_2 ) \Rightarrow |v\> = 0$. Again, the PWAK approach
is immune to these kind of arguments that, as Pauli's, require
boundedness of the Hamiltonian spectrum, which is not the case of the
clock Hamiltonian considered here.

While the basic mechanism to overcome the Pauli and Peres objections
presented here is clear, one has to be careful, since in the PWAK
formalism (reviewed in Sec.~\ref{s:paw}) the dynamics is imposed as a
constraint and one must check that even in the space of physical
states the above properties still hold true: indeed, in the space of
physical states, the Wheeler-de Witt equation ``forces'' the clock
Hamiltonian to coincide with the system Hamiltonian. This analysis is
given in Sec.~\ref{s:pauli} and \ref{s:reg} (that contains the more
technical parts). Even though the system Hamiltonian and the time
operator commute in this framework, it is still possible to give an
time-energy uncertainty relation, as shown in Sec.~\ref{s:cloc}.

\section{The PWAK mechanism\label{s:paw}}
The PWAK mechanism was initially proposed by Page and Wootters
\cite{paw,page,wootters} and soon after by Aharanov and Kaufherr
\cite{ak} (but similar previous approaches can be found, e.g.~in
\cite{morse,papersinvedral}). A recent review, together with the
solution to the objections that were moved against it, can be found in
\cite{nostro}.

To provide a quantization of time, one can simply define time as
``what is shown on a clock'' and then use a quantum system as a clock.
If one wants a continuous time that goes from $-\infty$ to $+\infty$,
a good candidate clock is to use the position of a 1-d particle
\cite{morse,ak,saleckerwigner}.  Nonetheless, introducing explicitly a
physical system is not necessary, and one can only consider the
time Hilbert space as an abstract space with no physical meaning. 

The global Hilbert space is then ${\cal H}_{TS}={\cal H}_T\otimes{\cal
  H}_S$, where $T$ represents the ``time'' Hilbert space, typically
the one for a particle on a line, ${\cal L}^2({\mathbb{R}})$. In
${\cal H}_T$ we introduce the position operator $\hat T$ and the
conjugate momentum $\hat\Omega$, with $[\hat T,\hat\Omega]=i$. We
associate the momentum $\hat\Omega$ to the energy of the clock (for a
particle, this can be a good approximation for sufficiently massive
non relativistic particles \cite{ak}). We can enforce that $\hat T$
represents the time operator which describes the evolution of a system
by imposing the following constraint equation, namely by requiring
that the only states $|\Psi\>\>$ of the joint Hilbert space ${\cal
  H}_{TS}$ that represent physically relevant situations are the ones
that satisfy
\begin{eqnarray}
(\hbar\hat\Omega\otimes\openone_S+\openone_{T} \otimes H)|\Psi\>\>=0
\labell{wdw}\;,
\end{eqnarray}
($H$ being the arbitrary Hamiltonian of the system $S$) which can be
interpreted as a Wheeler-de Witt equation \cite{wdw}. The double ket
notation serves only as a reminder that $|\Psi\>\>$ is a state on the
joint Hilbert space ${\cal H}_{TS}$. Note that, even though the system
Hamiltonian $H$ may have arbitrary spectrum, the Wheeler-de Witt
Hamiltonian $(\hbar\hat\Omega\otimes\openone_S+\openone_{T} \otimes
H)$ has unbounded continuous spectrum (because $\hat\Omega$ has).  As
eigenstates of the Wheeler-de Witt equation, the physical states
$|\Psi\>\>$ are ``static'' in the sense that they do not evolve with
respect to an ``external'' time. However, the system evolves with
respect to the clock and viceversa, in the sense that the correlations
(entanglement) between system and clock track the system evolution.
Indeed the solutions of \eqref{wdw} are 
\begin{eqnarray}
|\Psi\>\>=\int_{-\infty}^{+\infty}d\omega\;
|\omega\>|\tilde\psi(\omega)\>
\labell{solutions}\;,
\end{eqnarray}
where $|\omega\>$ is the eigenstate of $\hat\Omega$ with eigenvalue
$\omega$, $|\tilde\psi(\omega)\>$ is the (un-normalized) Fourier
transform of the system state $|\psi({t})\>$. [Note that the state
\eqref{solutions} is not uniform in $|\omega\>$, as the eventual
weight is implicit in the norm of $|\tilde\psi(\omega)\>$, e.g.~such
weight selects the solutions of \eqref{wdw}.] Indeed,
\begin{eqnarray}
  &&|\tilde\psi(\omega)\>=\tfrac
  1{\sqrt{2\pi}}\int_{-\infty}^{+\infty}dt\;e^{-i\omega {t}}|\psi({t})\>
  \Rightarrow\labell{ppp}
  \\&&\!\!\!\!\!
  H|\tilde\psi(\omega)\>=\sqrt{2\pi}\sum_k\delta(\omega_k+\omega)
  \psi_k\hbar\omega_k|e_k\>=
  -\hbar\omega|\tilde\psi(\omega)\>,\nonumber\\&&
\labell{psiom}
\end{eqnarray}
where we have used the expansion
$|\psi({t})\>=\sum_k\psi_ke^{-i\omega_kt}|e_k\>$ in terms of the
Hamiltonian eigenstates $|e_k\>$ of eigenvalue $\hbar\omega_k$. If the Hamiltonian has a
continuous spectrum, an analogous expression holds:
\begin{align}
  H|\tilde\psi(\omega)\>=\sqrt{2\pi}\int d\omega'\delta(\omega'+\omega)
  \psi(\omega')\hbar\omega'|\omega'\>
\nonumber\\
=
-\hbar\omega\sqrt{2\pi}\psi(-\omega)|-\omega\>=
  -\hbar\omega|\tilde\psi(\omega)\>,
\labell{cont}\;
\end{align}
where $|\omega\>$ is the $\delta$-normalized energy eigenstate of
eigenvalue $\hbar\omega$. It is clear from these expressions that
$|\tilde\psi(\omega)\>$ is the null vector if $\omega$ is not an
eigenvalue of the Hamiltonian. The solutions \eqref{solutions} can be
written as
\begin{eqnarray}
  |\Psi\>\>=\int_{-\infty}^{+\infty}d\omega\;
  |\omega\>|\tilde\psi(\omega)\>=
  \int_{-\infty}^{+\infty}dt\;|{t}\>|\psi({t})\>
\labell{solu}\;,
\end{eqnarray}
where $|{t}\>=\int d\omega\:e^{-i\omega {t}}|\omega\>/\sqrt{2\pi}$ is
the position eigenstate in ${\cal H}_T$ and $|\psi({t})\>$ is the
system state at time $t$ in ${\cal H}_S$, with normalization
$\<\psi({t})|\psi({t})\>=1$ for all $t$. [Note that any nontrivial
probability amplitude $\phi(\omega)$ in the integral \eqref{solutions}
can be absorbed in the definition of the system state
$|\tilde\psi(\omega)\>$ as $\psi_k\to\phi(\omega_k)\psi_k$.] The
states \eqref{solu} are improper (non-normalizable) states that reduce
to the momentum eigenstate $\sqrt{2\pi}|\omega=0\>=\int dt|{t}\>$ in
${\cal H}_T$ whenever the system is in an eigenstate of its
Hamiltonian $H$.  Starting from the state \eqref{solu} and
conditioning the clock system to the state $|{t}\>$, we recover the
Schr\"odinger equation: indeed, Eq.~\eqref{wdw} in the position
representation becomes
\begin{eqnarray}
\<{t}|\hbar\hat\Omega+H|\Psi\>\>=0\Leftrightarrow
(-i\hbar\tfrac\partial{\partial{t}}+H)|\psi({t})\>=0
\labell{schr}\;,
\end{eqnarray}
where we wrote the momentum in position representation
$\<{t}|\hat\Omega=(-i\partial/\partial {t})\<{t}|$, and we used
$\<{t}|\Psi\>\>=|\psi({t})\>$ which follows from \eqref{solu}. One can
similarly also derive the unitary evolution for the system
\cite{nostro}.

One of the many advantages of this approach is that it renders
explicit the problem that, when an event is gauged by a quantum clock
or a system is controlled by a quantum clock, a feedback (disturbance)
to the clock must occur \cite{peressaleckerwigner}.

\section{Bypassing the Pauli objection\label{s:pauli}}
The Pauli objection is just an argument and is not really rigorous.
There are many counterexamples in the literature (e.g.~\cite{busch}),
but it can also be made into a rigorous statement if one is careful
enough (e.g.~\cite{sri}\comment{This is probably not the most
  significant reference to put here!!}). It basically says that if one
introduces a time operator, then time and energy are conjugate
operators through the Schr\"odinger equation.  Then their spectrum
must be the same.  This is a consequence of the Stone-von Neumann
theorem: if $[\hat T,\hat H]=i$ then $\hat T$ and $\hat H$ have the
same spectrum.

The PWAK mechanism is immune to this, since we are requesting that
$[\hat T,\hat\Omega]=i$ and then enforcing that $\hat\Omega$ is equal
to $\hat H$ only on the physical states through the constraint
Eq.~\eqref{wdw}. Such equation is saying that in this subspace
$\hat\Omega=\hat H$! So it seems that in the space of physical states,
the Pauli argument should apply: $\hat T$ has the same spectrum as
$\hat\Omega$ which (in the subspace) has the same spectrum as $\hat
H$. So we must conclude that in the subspace of physical states $\hat
T$ has the same spectrum as $\hat H$!

Luckily this statement is false, although it is not immediately
trivial to see.  To see why that statement is false, we must formalize
it very carefully. We start by defining $\hat T$ and $\hat\Omega$ as
\begin{eqnarray}
\hat {T}&\equiv&\int_{-\infty}^{+\infty}dt\;{t}\;|{t}\>\<{t}|,\ 
\hat
{\Omega}\equiv\int_{-\infty}^{+\infty}d\omega\;{\omega}\;|{\omega}\>\<{\omega}|
\labell{tomega}\;
\\\nonumber
|\omega\>&\equiv&\int_{-\infty}^{+\infty}\tfrac{dt}{\sqrt{2\pi}}\;e^{-i\omega{t}}|{t}\>
,\
\Rightarrow|t\>=\int_{-\infty}^{+\infty}\tfrac{d\omega}{\sqrt{2\pi}}\;
e^{i\omega{t}}|{\omega}\>.
\end{eqnarray}
The Pauli objection can be formalized as follows:
\begin{enumerate}
\item The definitions of $\hat T$ and $\hat\Omega$ imply that $[\hat
  {T},\hat\Omega]=i$.
\item Introduce the Hilbert space of physical states ${\cal H}_{c}$ as
  the ones that satisfy Eq.~\eqref{wdw}, $(\hbar\hat\Omega+\hat
  H)|\Psi\>\>=0$.
\item\label{crit} Since $(\hbar\hat\Omega+\hat H)|\Psi\>\>=0$,
  then$^*$ also $\hat T(\hbar\hat\Omega+\hat H)|\Psi\>\>=0$, and
  $\<\<\Phi|\hat T(\hbar\hat\Omega+\hat H)|\Psi\>\>=0$ for all
  $|\Phi\>\>,|\Psi\>\>\in{\cal H}_c$.
\item The point above implies that \begin{eqnarray}
0&=&\<\<\Phi|\hat T(\hbar\hat\Omega+\hat
  H)|\Psi\>\>-\<\<\Phi|(\hbar\hat\Omega+\hat
  H)\hat T|\Psi\>\>\label{qq}\\&=&
\<\<\Phi|\hbar[\hat T,\hat\Omega]+[\hat T,\hat H]|\Psi\>\>,\\\mbox{so }&&
\<\<\Phi|\hbar[\hat T,\hat\Omega]|\Psi\>\>=-\<\<\Phi|[\hat T,\hat H]|\Psi\>\>
\labell{qui}\;
\end{eqnarray}
\item Since $[\hat {T},\hat\Omega]=i$, this means that, when
  restricting to the physical states space ${\cal H}_c$, we have
  $[\hat{T},\hat H]=-i\hbar$, which through the Stone-von Neumann
  theorem implies that $\hat T$ has the same spectrum as $\hat H$ in
  this Hilbert space ${\cal H}_c$, the Pauli objection!
\end{enumerate}

We note that Dirac had introduced an equation of the type \eqref{wdw}
in \cite{dirac}, but he did not consider it as a constraint on the
physical states. This meant that he ran into an inconsistency similar
to the one emphasized above. Dirac never gave a solution
\cite{hilgevoord}. We show here that a solution is provided by the
PWAK mechanism.

This above argument is clearly wrong since $[\hat{T},\hat H]=0$
because they are operators acting on different Hilbert spaces. In
fact, the implication indicated with an asterisk at point \ref{crit}
fails: even though it is true that $(\hbar\hat\Omega+\hat
H)|\Psi\>\>=0$, this does {\em not} imply that $\hat
T(\hbar\hat\Omega+\hat H)|\Psi\>\>=0$. This comes from the fact that
the spectrum of $\hat T$ is unbounded, see the definition
\eqref{tomega}. We prove this in the following section (using two
different regularizations for the physical states $|\Psi\>\>$ which
are un-normalizable).

One can also give a physical interpretation to this: one should expect
that the expectation value of $\hat T$ must be undefined in the space
of physical states. In fact $\<\hat{T}\>$ has as value the result to
the question ``what is the time?'' which is a meaningless question per
se in physics.  Meaningful questions are ``what is the time when the
spin is up?'' or ``what is the time now that you're reading this?'',
etc. So, one must expect that $\<\<\Phi|\hat T|\Psi\>\>$ will be
undefined in the space of physical states, which is indeed what
happens.
\section{Regularization\label{s:reg}}
\comment{conti in pauli-argument.xoj} Here we provide the
regularizations necessary to prove the relations introduced in
Sec.~\ref{s:pauli}.

The state \eqref{solu} is the solution of the eigenvalue equation
\eqref{wdw}. The eigenvalue $\lambda=0$ is an essential eigenvalue
of the self-adjoint constraint operator $\hat{\mathbb
  J}=\hbar\hat\Omega+\hat H$.  This can be shown through Weyl's
criterion \cite{ess} (Chap.~7), since $\|(\hat{\mathbb
  J}-\lambda)|\Psi_n\>\>\|\to 0$ for $n\to\infty$ where $|\Psi_n\>\>$
is a Weyl sequence, i.e.~a normalized sequence of vectors that weakly
converges to zero (namely, $\forall |\theta\>\>\in{\cal H}$ we have
$\<\<\theta|\Psi_n\>\>\to 0$).

We will show this using two different Weyl sequences which can be
considered as approximate eigenvectors (as expected, both give the same
results):
\begin{eqnarray}
|\Psi_n\>\>&\equiv&\left(\tfrac {2}{\pi n}\right)^{1/4}\int
dt\;e^{-t^2/n}|{t}\>|\psi({t})\>\\
|\Psi'_m\>\>&\equiv&\tfrac 1{\sqrt{m}}\int
dt\;\beta({t}/m)|{t}\>|\psi({t})\>
\labell{regu}\;,
\end{eqnarray}
where the first uses a Gaussian whose width diverges for $n\to\infty$,
the second uses the box function $\beta$ whose width diverges for
$m\to\infty$, with $\beta(x)=1$ if $-\tfrac12<x<\tfrac12$,
$\beta(-\tfrac12)=\beta(\tfrac12)=\tfrac12$, and $\beta(x)=0$
otherwise. It has derivative $\partial\beta(x)/\partial
x=\delta(x+1/2)-\delta(x-1/2)$. These are both Weyl sequences
(see \cite{ess} at pages 71 and 74 respectively).

All states in ${\cal H}_{c}$ can be obtained from these as
\begin{eqnarray} |\Psi\>\>=\lim_{n\to\infty}\left(\tfrac{\pi
      n}2\right)^{1/4}|\Psi_n\>\>=\lim_{m\to
    \infty}\sqrt{m}|\Psi'_m\>\>.
\labell{lim}\;
\end{eqnarray}
Note that the state $|\Psi\>\>$ is un-normalizable: it does not live
in a Hilbert space, but one has to resort to rigged Hilbert spaces,
where the Hilbert space containing normalized vectors (and the limit
of sequences of normalized vectors) is incremented with vectors of
infinite norm \cite{ballent}.

First we show that $|\Psi_n\>\>$ and $|\Psi'_m\>\>$ are indeed Weyl
sequences for $\hat{\mathbb J}$ and $\lambda=0$, namely they are
``proper'' approximations of the improper eigenvectors of
$\hat{\mathbb J}$ with eigenvalue $\lambda=0$. (We already know that,
this is just a consistency check.) Let us start with $|\Psi_n\>\>$. We
have to show that $\|(\hat{\mathbb J}-\lambda)|\Psi_n\>\>\|\to 0$ for
$n\to\infty$. Indeed,
\begin{eqnarray}
\lim_{n}(\hbar\hat\Omega+\hat H)|\Psi_n\>\>=
\lim_{n} \tfrac{2i}n\left(\tfrac{2}{\pi n}\right)^{1/4}\int
dt\;{t}\;e^{-t^2/n}
|{t}\>|\psi({t})\>.
\nonumber\labell{a1}\;
\end{eqnarray}
That this is a null vector can be seen by taking its modulus:
\begin{eqnarray}
\labell{l}\;
\|(\hbar\hat\Omega+\hat H)|\Psi_n\>\>\|^2
=\tfrac 4{n^2}\sqrt{\tfrac2{\pi n}}\int dt\;{t}^2\;e^{-2{t}^2/n}\to 0,\nonumber
\end{eqnarray}
where we used the fact that $\int
dt\;{t}^2\;e^{-at^2}=\sqrt{\pi/a^3}/2$ and
$\<\psi({t})|\psi({t})\>=1$.  The same result applies using the other
regularization: $(\hbar\hat\Omega+\hat H)|\Psi'_m\>\>\to 0$ for
$m\to\infty$. Indeed, for all vectors $|\theta\>\>=\int
dt\theta({t})|{t}\>|\phi({t})\>$ in the Hilbert space, we find
\begin{eqnarray}
&&|\<\<\theta|(\hbar\hat\Omega+\hat
H)|\Psi\>\>|\\&&\nonumber
=\lim_m|\<\<\theta|\int
dt\;|{t}\>|\psi({t})\>[\delta(t-\tfrac m2)-\delta(t+\tfrac m2)]
\\&&=\lim_m|\<\phi(0)|\psi(0)\>[\theta^*(\tfrac m2)-\theta^*(-\tfrac
m2)]|=0
\labell{ll}\;,
\end{eqnarray}
where we used \eqref{lim}, and the fact that $\<\phi({t})|\psi({t})\>$
is constant and that all square integrable functions $\theta({t})\to
0$ for ${t}\to\pm\infty$.

Now the crucial point: what happens when we multiply these null
vectors by the unbounded operator $\hat T$? We obtain a non-null
vector! Indeed, \begin{eqnarray}
\|\hat{T}(\hbar\hat\Omega+\hat H)|\Psi_n\>\>\|^2=
\tfrac 4{n^2}\sqrt{\tfrac2{\pi n}}\int dt\;{t}^4\;e^{-2{t}^2/n}=\tfrac34
\labell{asa}\;,\nonumber
\end{eqnarray}
for all $n$, since $\int dt\;{t}^4\;e^{-at^2}=3\sqrt{\pi/a^5}/4$. This
implies that $|\Psi_n\>\>$ is not an approximate eigenvector for the
$\lambda=0$ eigenvalue of the operator $\hat{T}(\hbar\hat\Omega+\hat
H)$, even though it was an approximate eigenvector for the operator
$(\hbar\hat\Omega+\hat H)$. This also means that in the rigged Hilbert
space we cannot consider $|\Psi\>\>$ as eigenvector of this operator.

One can also show that 
\begin{eqnarray}
&&\<\<\Psi_n|(\hbar\hat\Omega+\hat H)\hat{T}|\Psi_n\>\>
\\&&
=i\sqrt{\tfrac 2{\pi n}}\int
dt(e^{-2t^2/n}-\tfrac{2t^2}ne^{-2t^2/n})=\tfrac i2
\labell{las}\;,
\end{eqnarray}
which suggests that $|\Psi\>\>$ is an (improper) eigenstate of
$(\hbar\hat\Omega+\hat H)\hat{T}$ with eigenvalue $i/2$. Indeed, the above
results imply
\begin{eqnarray}
\|[(\hbar\hat\Omega+\hat H)\hat{T}-\lambda]|\Psi_n\>\>\|\to 0
\labell{lsad}\;
\end{eqnarray}
for $\lambda=i/2$ (actually the modulus is equal to 0 for all
$n$). Note that this is the value that is necessary in Eq.~\eqref{qq}
to avoid the contradiction!

Analogous considerations hold for the other regularization since
\begin{eqnarray}
\<\<\theta|\hat{T}(\hbar\hat\Omega+\hat H)|\Psi\>\>=
\lim_mi\tfrac{{m}}2[\theta^*(\tfrac m2)+\theta^*(-\tfrac m2)]
\labell{sk}\;,
\end{eqnarray}
where we used $|\Psi\>\>=\lim_m\sqrt{m}|\Psi'_m\>\>$. This does not
tend to zero as $m\to\infty$ for all square integrable functions
$\theta({t})$, since square integrable functions must go to zero
faster than $1/\sqrt{{t}}$ for $t\to\infty$.

In conclusion, not only we have shown that point \ref{crit} of the
Pauli argument fails, but we have also recovered the expected values
of the scalar products $\<\<\Psi|\hat{T}(\hbar\hat\Omega+\hat
H)|\Psi\>\>=i/2$ that are necessary in Eq.~\eqref{qq} if it has to be
consistent with the fact that $[\hat{T},\hat\Omega]=i$.

\section{Unbounded-energy clocks?\label{s:cloc}}
The way we the PWAK mechanism bypasses the Pauli objection is by using
a clock with an unbounded Hamiltonian equal to its momentum \cite{ak}.
Clearly this is unphysical and one could object that our resolution is
not a resolution after all. However, it is important to notice that
all quantum experiments to date have been performed with
macroscopic ``classical'' clocks (except for especially crafted
situations \cite{esperimentotorino}). These have energy so large
compared to the time uncertainties that can be tracked in practice
that their spectrum can be considered unbounded for all practical
purposes. Moreover, macroscopic systems get very quickly correlated to
astronomical distances (e.g.~the motion of one gram of matter on the
star Sirius by one meter sensibly influences the particle trajectories
in a box of gas on earth on a time-scale of $\mu$s after the transit
time~\cite{borel}) so that a pure-state analysis as performed above
will break down unless one is able to track all the correlated degrees
of freedom, a practical impossibility.

In this section we study how good is the approximation of considering
a clock with unbounded spectrum. We show that if the energy spread is
$\Delta E$ the time can be measured up to a precision $\Delta
{t}=\hbar/2\Delta E$. This is a direct consequence of the time-energy
uncertainty relation \cite{man,bata,qlimit} which says that, if the
energy spread is $\Delta E$, then the minimum time interval it takes
to evolve to an orthogonal state is $\tau\geqslant\hbar/2\Delta E$.
Hence no smaller time interval can be measured with accuracy.

Clearly, a spread in energy is by itself insufficient to obtain a
clock: one also needs good time correlation. However, in the absence
of energy spread, a clock in the state $|\Psi\>\>$ of
\eqref{solutions} cannot keep time, and with limited energy spread, it
can only keep time up to some accuracy since the correlation in time
cannot be sufficiently high. For the sake of simplicity we will keep
the unbounded Hamiltonian of the clock $\hbar\Omega$, but we will
reduce the energy spread by making explicit the spectral function
$\phi(\omega)$ which was absorbed into $|\tilde\psi(\omega)\>$ in
\eqref{solutions} as
$|\tilde\psi(\omega)\>=\phi(\omega)|\tilde\chi(\omega)\>$, and we take the
standard deviation of the probability probability $|\phi(\omega)|^2$
is $\Delta\omega=\Delta E/\hbar$.  Since the clock and the system are
entangled, this spectral function {\em does not refer exclusively to
  the clock}, but to both the clock and the system. As expected from
the time-energy uncertainty relation, a limited-bandwidth spectral
function $\phi(\omega)$ will reduce the speed of evolution (time
resolution of the global system).  Indeed (neglecting multiplicative
constants) we have
\begin{eqnarray}
  |\Psi\>\>=\int d\omega\:\phi(\omega)
  |\omega\>|\tilde\chi(\omega)\>\propto
  \int dtdt'\:\tilde\phi({t}-t')|{t}\>|\chi(t')\>
  \nonumber,
\end{eqnarray}
where $\tilde\phi$ and $|\chi({t})\>$ are the Fourier transforms of
$\phi$ and $|\tilde\chi(\omega)\>$. Even though this seems to be
incompatible with Eq.~\eqref{solu}, it is not as can be seen by
writing 
\begin{align}
|\psi({t})\>\propto\int dt'\tilde\phi({t}-t')|\chi(t')\>
\labell{aver}\;.
\end{align}
This can be interpreted as if $|\psi({t})\>$ is obtained by
``averaging'' $|\chi({t})\>$ over time with a probability amplitude
$\tilde\phi$. Then the smallest time interval during which
$|\psi({t})\>$ can vary appreciatively is of the order of
$\hbar/\Delta E$, i.e.~the inverse of the spread of the probability
$|\phi(\omega)|^2$. Indeed
\begin{align}
\<\psi({t})|\psi(t')\>\propto\int d\tau d\tau'\;
\tilde\phi({t}-\tau)
\tilde\phi^*({t'}-\tau')\<\chi(\tau')|\chi(\tau')\>
\nonumber,
\end{align}
whence, even supposing instantaneous change of $|\chi\>$, namely
$\<\chi(\tau')|\chi(\tau')\>\propto\delta(\tau-\tau')$, we have 
\begin{align}
\<\psi({t})|\psi(t')\>\propto\int d\omega
|\phi(\omega)|^2e^{i\omega({t}-t')}
\label{qqq}.
\end{align}
If $|\phi(\omega)|^2$ has a spread $\simeq\Delta E/\hbar$, then its
Fourier transform will have a spread of the order of $\hbar/\Delta E$.
This means that the scalar product $\<\psi({t})|\psi(t')\>$ cannot
change appreciatively in a smaller interval, namely the time scale of
change of the system state must be larger than $\hbar/\Delta E$, in
accordance with the time-energy uncertainty relation.

Thanks to its linearity, the Schr\"odinger equation holds for the
``averaged'' $|\psi({t})\>$ of Eq.~\eqref{aver}, which implies that
eventual imperfect correlations between system and clock will not
induce a fundamental decoherence effect in this framework.  This
contrasts to the Gambini et al.  framework
\cite{gambinipullin,montevideo} where imperfect clocks do induce a
fundamental decoherence.

\section{Conclusions\label{s:concl}}
In this paper we have shown how one can easily bypass the Pauli and
the Peres objections to a quantum operator for time using the
conditional probability amplitude framework of Page, Wootters,
Aharanov and Kaufherr. Moreover we have detailed how the time-energy
uncertainty relation arises in this framework.

\section*{Acknowledgments}
We acknowledge the FQXi foundation for financial support in the
``Physics of what happens'' program. JL acknowledges support from
MINECO/FEDER Project FIS2015-70856-P and CAM PRICYT Project QUITEMAD+
S2013/ICE-2801 and Giacomo Mauro D'Ariano for the kind hospitality at
Pavia University. LM acknowledges very useful feedback from Vittorio
Giovannetti and Seth Lloyd.


\begin{references}
\bibitem{kuchar}K.V. Kucha\u r, ``Time and interpretations of quantum
  gravity'', Proc. 4th Canadian Conference on General Relativity and
  Relativistic Astrophysics, eds. G. Kunstatter, D.  Vincent, and J.
  Williams (World Scientific, Singapore, 1992), pg.~69-76.
\bibitem{anderson}E. Anderson, ``The Problem of Time in  Quantum Gravity"
  in {\em Classical and Quantum Gravity: Theory, Analysis and Applications}, Ed. V.R. Frignanni.Nova, New York 2012.
\bibitem{problemoftime} K.V. Kucha\u r, in {\em Quantum Gravity 2: a Second
  Oxford Symposium}, ed. C.J. Isham, R. Penrose and D.W. Sciama
  (Clarendon, Oxford 1981); K.V. Kucha\u r, in {\em Conceptual Problems of
  Quantum Gravity}, ed. A. Ashtekar and J. Stachel (Birkh\" auser,
  Boston 1991); C.J. Isham, in {\em Integrable Systems, Quantum Groups and
  Quantum Field Theories}, ed. L.A. Ibort and M.A. Rodr\' iguez (Kluwer,
  Dordrecht 1993); K.V. Kucha\u r, in {\em The Arguments of
  Time}, ed. J. Butterfield (Oxford University Press, Oxford 1999).
\bibitem{popescu}Y. Aharonov, J. Oppenheim, S. Popescu,  B. Reznik,
  W. G. Unruh, ``Measurement of time of arrival in quantum mechanics", Phys. Rev. A {\bf 57}, 4130 (1998).
\bibitem{werner}R. Werner, ``Arrival time observables in quantum mechanics",  Annales de l'I. H. P., section A, {\bf 47},
  429 (1987).
\bibitem{mielnik}B. Mielnik,``The Screen Problem", Found. Phys. {\bf 24}, 1113 (1994).
\bibitem{delgado}V. Delgado, J. G. Muga, ``Arrival time in quantum mechanics",  Phys. Rev. A {\bf 56,} 3425
  (1997).
\bibitem{galap}E.A. Galapon, A. Villanueva, ``Quantum first time-of-arrival operators", J. Phys. A: Math. Theor.  {\bf 41}, 455302 (2008).
\bibitem{juan}J. Leon, ``Time-of-arrival formalism for the
  relativistic particle'', J. Phys. A: Math. Gen. {\bf 30}, 479 (1997).
\bibitem{nw}T. D. Newton, E. P. Wigner, ``Localized States for
  Elementary Systems'', Rev. Mod. Phys. {\bf 21}, 400 (1949).
\bibitem{freden}R. Brunetti, K. Fredenhagen, M. Hoge, ``Time in quantum physics: From an external parameter to an intrinsic observable ", Found. Phys.  {\bf 40}, 1368 (2010);  R. Brunetti, K. Fredenhagen,  ``Time of
 occurrence observable in quantum mechanics'',   Phys. Rev. A {\bf 66}, 044101 (2002).
\bibitem{olkhovski}V. S. Olkhovsky, E. Recami,``Time as a Quantum Observable", Int. J. Mod. Phys. A   {\bf 22}, 5063 (2007).
\bibitem{olkhovski1}V.S. Olkhovsky,``Time as a Quantum Observable, Canonically Conjugated to Energy, and Foundations of Self-Consistent Time Analysis of Quantum Processes", Adv. Math. Phys. {\bf 2009},
  859710 (2009).
\bibitem{caves} T. G. Downes, G. J. Milburn, C. M. Caves,``Optimal Quantum Estimation for Gravitation",   arXiv:1108.5220 (2011).
\bibitem{stu} E.C.G. Stueckelberg, `` La signification du temps
    propre en m\'ecanique ondulatoire", Helv. Phys. Acta {\bf 14,} 322
  (1941); E.C.G. Stueckelberg, `` La m\'ecanique du point mat\'eriel
    en th\'eorie des quanta", Helv. Phys. Acta {\bf 15,} 23 (1942).
\bibitem{fanchi}J. R. Fanchi, ``Review of invariant time formulations of relativistic quantum theories", Found. Phys. {\bf 41}, 4 (2011).
\bibitem{ahar} Y. Aharonov, S. Popescu, J. Tollaksen, ``Each instant of
time a new Universe", 
  in {\em Quantum Theory: A Two-Time Success Story} (Springer, 2014)
  pg. 21-36, arXiv:1305.1615 (2013).
\bibitem{farhi}E. Farhi, S. Gutmann, ``The functional integral constructed  directly from the Hamiltonian", Ann. Phys. {\bf 213}, 182 (1992).
\bibitem{wilcz} J. Cotler, F. Wilczek, ``Entangled histories'',
  Physica Scripta {\bf T168}, 014004 (2016).
  % Entangled History, arXiv:1502.02480 (2015).
\bibitem{peresbook}A. Peres, {\em Quantum Theory: Concepts and
    Methods} (Kluwer ac. publ., Dordrecht, 1993).
%\bibitem{darwinism}W.H. Zurek, %Quantum Darwinism,
%  Nature Phys. {\bf 5}, 181 (2009).
\bibitem{pauli}W. Pauli, {\em General Principles of Quantum Mechanics}
  (Springer, Berlin Heidelberg, 1980).
\bibitem{milburn}S. L. Braunstein, C. M. Caves, G. J. Milburn,``Generalized Uncertainty Relations: Theory, Examples, and Lorentz Invariance", Annals   Phys. {\bf 247}, 135 (1996).
\bibitem{qspeed} V. Giovannetti, S. Lloyd, and L. Maccone,
``Quantum limits to dynamical evolution'', 
  Phys. Rev. A {\bf 67}, 052109 (2003).
\bibitem{holevo} A.S. Holevo, {\em Quantum Systems, Channels,
    Information} (de Gruyter Studies in Mathematical Physics).
\bibitem{holevo1} A.S. Holevo,``Estimation of shift parameters of a quantum state", Rep. Math. Phys. {\bf 13}, 379 (1978).
\bibitem{greenbergerproper}D. M. Greenberger,``Conceptual Problems Related to Time and Mass in Quantum Theory", arXiv:1011.3709 (2010).
\bibitem{saleckerwigner} H. Salecker, E.P. Wigner, ``Quantum
Limitations of the Measurement of Space-Time Distances", 
  Phys. Rev. {\bf 109}, 571 (1958).
\bibitem{peressaleckerwigner}A. Peres, ``Measurement of time by quantum clocks", Am. J. Phys. {\bf 48}, 552
  (1980).
\bibitem{paw}D.N. Page and W.K. Wootters, ``Evolution without evolution: Dynamics described by stationary observables", Phys. Rev. D, 27, 2885 (1983).
\bibitem{page}D.N. Page, ``Clock time and entropy'',
  in {\em Physical Origins of Time Asymmetry}, eds. J.J. Halliwell, et
  al., (Cambridge Univ. Press, 1993), arXiv:gr-qc/9303020.
\bibitem{wootters} W.K. Wootters, `` `Time' replaced by quantum correlations", Int. J. Theor. Phys. 23, 701 (1984).
\bibitem{ak}Y. Aharonov, T. Kaufherr, ``Quantum frames of reference", Phys. Rev. D {\bf 30}, 368    (1984).
\bibitem{nostro}V.Giovannetti, S.Lloyd, L.Maccone, ``Quantum time", Phys. Rev. D,  {\bf 92}, 045033 (2015).
\bibitem{morse}P. McCord Morse, H. Feshbach, {\em Methods of
    Theoretical Physics, Part I} (McGraw-Hill, 1953), Chap. 2.6.
\bibitem{zeht}H.D. Zeh,``Emergence of classical time from a universal wavefunction", Phys. Lett. A {\bf 116}, 9 (1986); H.D. Zeh,
  {\em Time in quantum theory},
  http://www.rzuser.uni-heidelberg.de/$\sim$as3/TimeInQT.pdf
\bibitem{vedral} V. Vedral,  ``Time, (Inverse) Temperature and
 Cosmological Inflation as Entanglement", 
  arXiv:1408.6965 (2014).
\bibitem{papersinvedral} T. Banks, ``TCP, quantum gravity, the cosmological constant and all that...", Nucl. Phys. B {\bf 249}, 332
  (1985); R. Brout, ``On the concept of time and the origin of the cosmological temperature", Found. Phys. {\bf 17}, 603 (1987); R. Brout,
  G. Horwitz,  D. Weil, ``On the onset of time and temperature in cosmology",  Phys. Lett. B {\bf 192}, 318 (1987); R. Brout, ``Time and temperature in semi-classical gravity",   Z. Phys. B {\bf 68}, 339 (1987).
\bibitem{rovelli} C. Rovelli, ``Relational Quantum Mechanics",
  Int. J. of Theor. Phys. {\bf 35}, 1637 (1996).
\bibitem{rovt}C. Rovelli, ``Time in quantum gravity: An hypothesis", Phys. Rev. D {\bf 43}, 442 (1991).
\bibitem{rovellibook}C. Rovelli {\em Quantum Gravity} (2003),
  obtainable from http://www.cpt.univ-mrs.fr/~rovelli/book.pdf.
%\bibitem{notata} This is reasonable from two points of views: (i)~from
%  the point of view of canonical general relativity, this is basically
%  equivalent to the Wheeler-De Witt equation; (ii)~from a Machian
%  perspective, it makes no sense to say that a shift of the whole
%  universe (including all the physical clocks) in time to be
%  measurable.
\bibitem{gambinipullin}R. Gambini, R.A. Porto, J. Pullin, S.
  Torterolo, ``Conditional probabilities with Dirac observables and
the problem of time in quantum gravity'', 
  Phys. Rev. D {\bf 79}, 041501(R) (2009).
\bibitem{montevideo} R. Gambini, L. P. Garcia-Pintos, J. Pullin, ``An
axiomatic formulation of the Montevideo interpretation of quantum
mechanics",   Studies in the History and Philosophy of Modern Physics, {\bf 42},  256 (2011); R. Gambini, J. Pullin, 
  `` The Montevideo interpretation of quantum mechanics: frequently
  asked questions", J. Phys. Conf. Ser. 174:012003, 2009.
\bibitem{esperimentotorino} E. Moreva, G. Brida, M. Gramegna, V.
  Giovannetti, L. Maccone, M. Genovese, ``Time from quantum
entanglement: an experimental illustration", 
  Phys. Rev. A {\bf 89}, 052122 (2014).
\bibitem{hilgevoord}J. Hilgevoord, ``Time in quantum mechanics: a story of confusion", Studies in History and Philosophy of   Modern Physics {\bf 36}, 29 (2005). 
\bibitem{wdw} B.S.  DeWitt, ``Quantum Theory of Gravity. I. The
  Canonical Theory''.  Phys. Rev. {\bf 160}, 1113 (1967).
\bibitem{peresobj} See \cite{peresbook}, Eq.~(8.90).
\bibitem{halvorson}H. Halvorson, ``Does quantum theory kill time?'',
  from http://www.princeton.edu/~hhalvors/papers/, (2010).
\bibitem{hegerfeldt} G. C. Hegerteldt and S. N. M. Ruijsenaars, ``Remarks on causality, localization, and spreading of wave packets", Phys. Rev D 22, 377 (1980).
\bibitem{busch}P. Busch, M. Grabowski, P.J. Lahti,``Time observables in quantum theory", Phys. Lett. A {\bf  191}, 357 (1994).
\bibitem{sri}M.D. Srinivas, R. Vijayalakshmi, ``The `time of occurrence' in quantum mechanics", Pramana {\bf 16}, 173   (1981).
\bibitem{dirac}P.A.M.
  Dirac, `` Relativity quantum mechanics with an application to Compton scattering".   Proc. Royal Soc. (London) A, {\bf 111}, 405 (1926).
\bibitem{ess} P.D. Hislop, I.M. Sigal, {\em Introduction to Spectral
    Theory}, Applied Mathematical Sciences {\bf 113} (Springer, 1996).
\bibitem{ballent}L.E. Ballentine {\em Quantum Mechanics, a Modern
    Development} (World Scientific, 2014), Sec. 1.4.
\bibitem{borel}E. Borel, {\em Le Hasard} (Alcan, Paris, 1914).
\bibitem{man} L. Mandelstam and I. G. Tamm, ``The Uncertainty Relation Between Energy and Time in Non-relativistic Quantum Mechanics", J. Phys. USSR {\bf 9}, 249 (1945).
\bibitem{bata}  K. Bhattacharyya,``Quantum decay and the Mandelstam -Tamm-energy inequality", J. Phys. A {\bf 16}, 2993 (1983).
\bibitem{qlimit} V. Giovannetti, S. Lloyd, and L. Maccone, {``Quantum
    limits to dynamical evolution'',} Phys. Rev. A {\bf 67}, 052109
  (2003).
\end{references}
\end{document}